\documentclass[10pt,twocolumn,letterpaper]{article}

\usepackage[cvpr]{cvpr}      

\usepackage{graphicx}
\usepackage{amsmath}
\usepackage{amssymb}
\usepackage{booktabs}

\usepackage{mwe} 

\usepackage{algorithm}
\usepackage{algpseudocode}
\usepackage{multirow}
\usepackage{xcolor}

\usepackage[pagebackref,breaklinks,colorlinks]{hyperref}

\usepackage[capitalize]{cleveref}
\crefname{section}{Sec.}{Secs.}
\Crefname{section}{Section}{Sections}
\Crefname{table}{Table}{Tables}
\crefname{table}{Tab.}{Tabs.}

\def\cvprPaperID{*****} 
\def\confName{CVPR}
\def\confYear{2022}

\begin{document}

\newcommand{\techname}{LegoPET}
\newcommand{\denet}{PnPNet}

\title{LegoPET: Hierarchical Feature Guided Conditional Diffusion for PET Image Reconstruction}

\author{Yiran Sun\\
Rice University\\
6100 Main St, Houston, TX 77005\\
{\tt\small ys92@rice.edu}
\and
Osama Mawlawi\\
The University of Texas MD Anderson Cancer Center\\
1155 Pressler St, Houston, TX 77030\\
{\tt\small omawlawi@mdanderson.org}
}

\maketitle

\begin{abstract}
Positron emission tomography (PET) is widely utilized for cancer detection due to its ability to visualize functional and biological processes \textit{in vivo}. PET images are usually reconstructed from histogrammed raw data (sinograms) using traditional iterative techniques (e.g., OSEM, MLEM). Recently, deep learning (DL) methods have shown promise by directly mapping raw sinogram data to PET images. However, DL approaches that are regression-based or GAN-based often produce overly smoothed images or introduce various artifacts respectively. Image-conditioned diffusion probabilistic models (cDPMs) are another class of likelihood-based DL techniques capable of generating highly realistic and controllable images. While cDPMs have notable strengths, they still face challenges such as maintain correspondence and consistency between input and output images when they are from different domains (e.g., sinogram vs. image domain) as well as slow convergence rates. To address these limitations, we introduce \techname{}, a hierarchica\textbf{L} f\textbf{e}ature \textbf{g}uided conditional diffusi\textbf{o}n model for high-perceptual quality PET image reconstruction from sinograms. We conducted several experiments demonstrating  that \techname{} not only improves the performance of cDPMs but also surpasses recent DL-based PET image reconstruction techniques in terms of visual quality and pixel-level PSNR/SSIM metrics. Our code is available at \href{https://github.com/yransun/LegoPET}{https://github.com/yransun/LegoPET}.
\end{abstract}

\section{Introduction}
\label{sec:intro}

Positron emission tomography (PET) is a functional imaging technique that visualizes various biochemical and physiological processes across different tissues. During the image acquisition process, high-energy photons generated by positron annihilation are detected and counted by detectors arranged in rings within a PET scanner. Usually, the measured counts from various angles form projections, which are then stacked into sinograms. These sinograms serve as the basis for reconstructing the final PET image, enabling the visualization of functional processes within the body \cite{fahey2002data}.

Conventional PET image reconstruction techniques, such as Ordered Subsets Expectation Maximization (OSEM) and Maximum Likelihood Expectation Maximization (MLEM), often encounter challenges like data-model mismatch, data inconsistency, and overfitting, which may introduce artifacts and noise into the reconstructed images \cite{haggstrom2019deeppet}. Recently, deep learning (DL) methods that directly transform raw PET sinograms into images have gained attention due to their ability to learn complex, non-linear physical processes from data. However, regression-based DL approaches are typically supervised with mean square error (MSE) or mean absolute error (MAE), leading to resultant images with overly smooth textures and suboptimal perceptual quality~\cite{ren2023multiscale,ledig2017photo}. Generative models, such as generative adversarial networks (GANs) \cite{goodfellow2020generative}, represent an advanced class of DL techniques designed to capture inherent data distributions and generate realistic images. Nevertheless, the underlying competitive nature between generator and discriminator segments of GAN models often leads to issues such as non-convergence and mode collapse during training \cite{salimans2016improved}. Image-conditioned diffusion probabilistic models (cDPMs) \cite{ho2020denoising, saharia2022palette, ho2022classifier} have lately demonstrated competitive performance, which exhibits better convergence behavior and generates more realistic images. However, cDPMs also face several key challenges: (1) insufficient correspondence and consistency between conditioning inputs and outputs, especially when they originate from different domains (e.g., sinogram vs. image domains), (2) difficulty in capturing high frequency details, and (3) low training efficiency \cite{han2023contrastive, zhang2024improving}.

To address above limitations, we propose \techname{}, a hierarchica\textbf{L} f\textbf{e}ature \textbf{g}uided conditional diffusi\textbf{o}n model for high-perceptual quality PET image reconstruction from sinograms. Specifically, building on previous work~\cite{ren2023multiscale, zhang2023adding, sun2024diffpet}, we first train a convolutional U-Net on sinogram-PET pairs as a plug-and-play prior (\denet{}), and then efficiently fine-tune a standard cDPM by using structured multilevel features from the learned \denet{} as biases. Additionally, we use the classifier-free training strategy to balance mode coverage and sample fidelity~\cite{ho2022classifier}. Experimental results demonstrate that \techname{} not only improves the performance of cDPMs but also surpasses recent DL-based PET image reconstruction techniques in terms of visual quality and pixel-level PSNR/SSIM metrics.

\section{Method}
\label{sec:format}
Let $\mathbf{s}$ denote input sinogram image, where $\mathbf{s} \in \mathbb{R}^{1 \times H \times W}$ is a single-channel 2D image with resolution $H \times W$. We denote the reference PET scan by $\mathbf{x}_0 \in \mathbb{R}^{1 \times H \times W}$. Our general objective is to approximate and sample from the conditional distribution $p(\mathbf{x}_0|\mathbf{s})$. We develop \techname{} to address this task, containing two components (see Fig.~\ref{fig:overview}): a learned plug-and-play prior network (right) that generates hierarchical feature maps, and a standard 2D sinogram-conditioned diffusion model (left) that reconstructs final clean PET image. We describe details of \techname{} in the following sections.

\subsection{Plug-and-Play Prior Network (\denet{})}
\begin{figure}[t!]
    \centering
    \includegraphics[width=\linewidth]{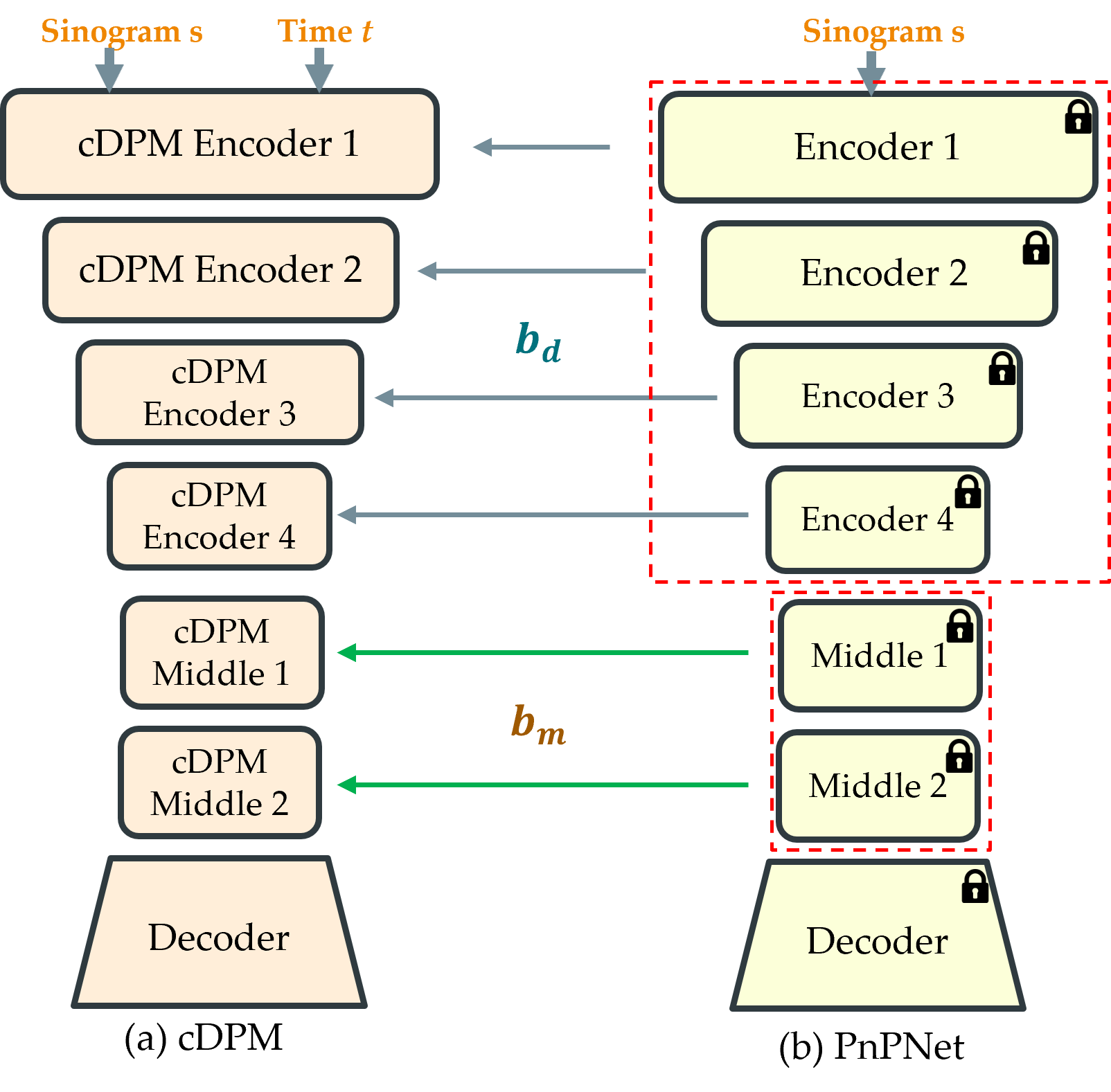}
    \caption{\textbf{Overview of \techname{}.} \techname{} includes two parts: a learned \denet{} for hierarchical feature maps extraction (right), and a standard 2D sinogram-conditioned DPM for PET image reconstruction (left). Two lists of structured layered feature maps, $b_d$ and $b_m$, from the pre-trained \denet{} are added to the cDPM as biases.}
\label{fig:overview}
\end{figure}

We utilize a conditional convolutional U-Net, which is the same architecture as commonly-used U-Net backbone in diffusion models, consisting of ResNet blocks \cite{he2016deep} and spatial self-attention blocks. Specifically, \denet{} comprises 4 encoder blocks, 2 middle blocks, and 4 decoder blocks, taking a single sinogram $\mathbf{s}$ as input and getting the corresponding PET image information (see Fig.~\ref{fig:overview}b). In our experiments, input and output have been padded with dimensions of $1 \times 256 \times 256$. During downsampling stage, the blocks operate at 4 resolutions: $256 \times 256$, $128 \times 128$, $64 \times 64$, $32 \times 32$. Note that \denet{}'s architecture is not limited to this design and may perform similarly well with other architectural variations.

\denet{} is trained in an end-to-end manner using spatial MSE-based ($\mathcal{L}_{MSE}$) and Discrete Wavelet Transform-based ($\mathcal{L}_{DWT}$) \cite{othman2020applications} losses for supervision on sinogram-PET pairs:
\begin{equation}
    \mathcal{L}_{\mathbf{\denet{}}} = \mathcal{L}_{MSE} + \lambda_{1} * \mathcal{L}_{DWT}
\end{equation}
where $\lambda_1$ controls the importance of DWT loss term (we set $\lambda_1=0.1$ in our experiments). $\mathcal{L}_{DWT}$ mainly focuses on calculating the difference between ground truth and prediction's high-frequency spectrum, which ensures that the hierarchical feature maps from \denet{} are able to additionally capture more high-frequency information (e.g. tissue edges, anatomical structures, textures).

Then we extract two lists of hierarchical feature maps from downsampling blocks ($b_d$) and middle blocks ($b_m$) in the latent space of learned \denet{}. We connect the pre-trained \denet{} to cDPM by incorporating $b_d$ and $b_m$ as extra biases to the encoder and middle blocks of cDPM. This integration allows the cDPM to leverage the low- and high-frequency information from learned feature representations.

\subsection{\techname{} for Sinogram-to-PET}
Conditional DPMs (cDPMs) are designed to learn a parameterized Markov chain that transforms a Gaussian distribution into a conditional data distribution. In particular, \techname{} approximates the conditional distribution through a fixed forward process and a learning-based reverse process.

The forward process starts with a clean PET sample from the input data distribution $\mathbf{x}_0 \sim q(\mathbf{x}_0)$ and gradually adds Gaussian noise according to a variance schedule $\beta_{1:T}$, where $\beta_t \in (0, 1)$ for all $t \in [1, T]$:
\begin{equation}
    q(\mathbf{x}_t | \mathbf{x}_{t-1}) := \mathcal{N}(\mathbf{x}_t;\sqrt{1-\beta_t}\mathbf{x}_{t-1}, \beta_t\textbf{I})
    \label{eq:diff1}
\end{equation}
where $\mathbf{x}_T$ is an isotropic Gaussian distribution for large enough $T$. We express $\mathbf{x}_t$ in closed form with respect to $\mathbf{x}_0$ directly, which allows for efficient training. Let $\alpha_t := 1-\beta_t$, $\bar{\alpha}_t := \prod^t_{s=1}\alpha_s$. Then we can sample $\mathbf{x}_t$ at any time step $t$ using a linear combination of noise $\epsilon \sim \mathcal{N}(0, \textbf{I})$ and $\mathbf{x}_0$:
\begin{equation}
    \mathbf{x}_t = \sqrt{\overline{\alpha}_t}\mathbf{x}_0 + \sqrt{1-\overline{\alpha}_t}\epsilon
    \label{eq:diff3}
\end{equation}

The goal of reverse process is to generate a clean PET image $\mathbf{x}_0$ from the noisy vector $\mathbf{x}_T$ given the conditions $\mathcal{C}$ = \{$\mathbf{s}, b_d, b_m$\}, i.e. approximate specific conditional distribution $p(\mathbf{x}_0 |\mathcal{C})$. We use a joint Markov chain distribution to model this process: $p_{\theta}(\mathbf{x}_{0:T}) := p(\mathbf{x}_T)\prod_{t=1}^T p_{\theta}(\mathbf{x}_{t-1} | \mathbf{x}_t, \mathcal{C})$, where $p(\mathbf{x}_T) = \mathcal{N}(\mathbf{x}_T; \textbf{0}, \textbf{I})$. We learn the transition $p_\theta(\mathbf{x}_{t-1}|\mathbf{x}_{t}, \mathcal{C})$ using a neural network $\mu_{\theta}(\cdot, \cdot)$:
\begin{equation}
    p_{\theta}(\mathbf{x}_{t-1} | \mathbf{x}_t, \mathcal{C}) := \mathcal{N}(\mathbf{x}_{t-1}; \mu_{\theta}(\mathbf{x}_t, t, \mathcal{C}), \textstyle\sum\nolimits_{\theta}(\mathbf{x}_t, t, \mathcal{C}))
\end{equation}
where $\theta$ represents the learnable parameters of the neural network. We can further reparameterize $\mu_{\theta}(\cdot, \cdot)$ by:
\begin{equation}
    \mu_{\theta}(\mathbf{x}_t, t, \mathcal{C}) = \frac{1}{\sqrt{\alpha_t}}\left(\mathbf{x}_t - \frac{1-\alpha_t}{\sqrt{1-\bar{\alpha}_t}}\epsilon_{\theta}(\mathbf{x}_t, t, \mathcal{C})\right)
\end{equation}
where $\epsilon_{\theta}(\cdot, \cdot)$ predicts the noise added at each time step.

\subsection{Training and Inference}
Following previous diffusion model-based work, we use a time-conditioned convolutional U-Net to perform denoising at each time step of the reverse process. We incorporate conditions $\mathcal{C}$ into the denoising process. During each iteration of training, a batch of random sinogram-PET pairs is selected, and \techname{} is trained by minimizing the denoising loss:
\begin{equation}
\mathcal{L}_{\mathbf{\techname{}}} := \mathbb{E}_{t\sim [1,T],\mathbf{x}_0,\epsilon}\left[||\epsilon_t - \epsilon_{\theta}(\mathbf{x}_t,t,\mathcal{C})||^2\right]
\end{equation}

We use classifier-free diffusion guidance training strategy. The conditioning information $\mathcal{C}$ is randomly removed with probability $p_{dp}$ and replaced with the same shape null tensors representing the absence of conditioning information. This strategy assists \techname{} to effectively capture both conditional and unconditional distributions, also the differences between them, thereby generating results that are more accurately aligned with the conditioning information \cite{ho2022classifier}.

At inference time, given a sample of Gaussian noise \(\mathbf{x}_T \sim \mathcal{N}(0, \mathbf{I})\), we use new denoising expression $\tilde{\boldsymbol{\epsilon}}_\theta = (1 + \lambda_{2}) \boldsymbol{\epsilon}_\theta (\mathbf{x}_t, t, \mathcal{C}) - \lambda_{2} \boldsymbol{\epsilon}_\theta (\mathbf{x}_t, t)$ to progressively denoise \(\mathbf{x}_T\) over \(T\) steps to generate a clean PET image \(\mathbf{x}_0\), where $\lambda_{2}$ controls the relative importance of different terms.

\section{Experiments}
\label{sec:pagestyle}


\begin{figure*}[t!]
    \centering
    \includegraphics[width=\textwidth]{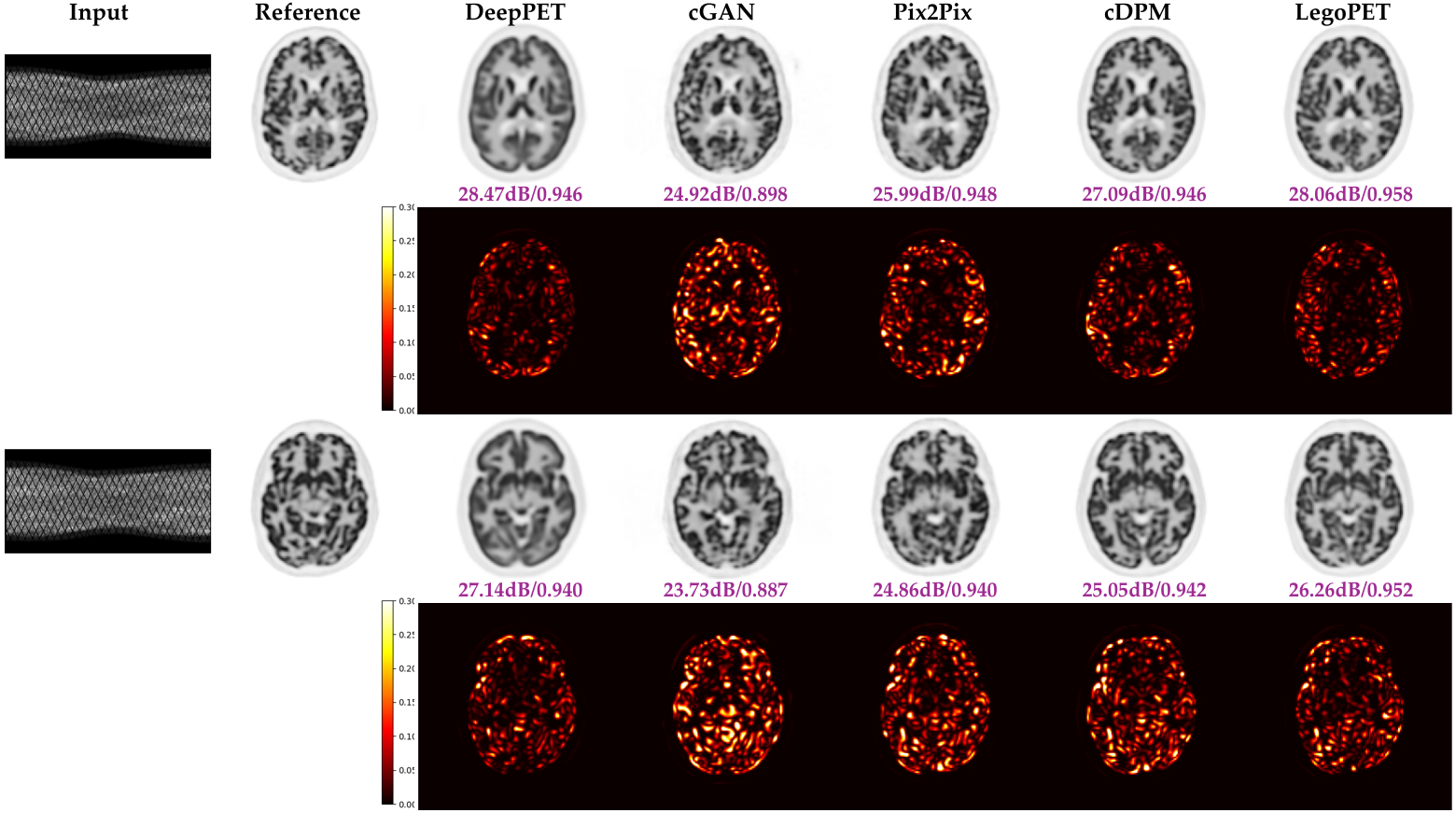}
    \caption{\textbf{Comparison of \techname{} with Four Baselines on Two Example Reconstructed Slices.} The first column shows the input sinogram images, and the second column shows the reference images reconstructed using OSEM algorithm. The third to sixth columns correspond to the four baselines (labeled above each image), and the final column shows the reconstructed PET image using proposed \techname{} method. PSNR/SSIM values are reported below each slice, and squared error maps between each method and the reference image are also displayed (second and fourth rows). We prove that \techname{} generates PET images with the highest perceptual quality, and also improves the performance of cDPM through feature guidance.}
    \label{fig:method}
\end{figure*}

\subsection{Datasets and Preprocessing}
We simulated 2D $^{18}$F-FDG PET images using the public 20 3D brain phantoms from BrainWeb \cite{collins1998design} with the resolution and matrix size of $2.086 \times 2.086 \times 2.031$ mm$^3$ and $344 \times 344 \times 127$ acquired from a Siemens Biograph mMR. We used data augmentation technique by rotating each 3D brain phantom 5 times. For each 3D brain phantom, we selected 55 non-continuous slices from axial view to generate high count sinograms which were used to reconstruct the reference PET images. We used 17 brain samples (4675 slices) for training, 1 brain sample (275 slices) for validation and 2 brain samples (550 slices) for testing.

\subsection{Implementation Details}
We implemented all experiments using PyTorch~\cite{paszke2019pytorch} on NVIDIA A100 GPUs. We trained \denet{} using the Adam optimizer with a fixed learning rate of $3 \times 10^{-5}$, and selected the model at epoch 119. We set batch size to 4 and trained 500 epochs for each \techname{} model, with $T=1000$ timesteps and $p_{dp} = \{0, 0.1, 0.2, 0.5\}$. We chose the final checkpoints for all \techname{}s without any specific selection strategies. We used pixel-level metrics for quantitative evaluation, i.e. Peak Signal to Noise Ratio (PSNR) and Structural Similarity Index Measure (SSIM) \cite{hore2010image}. Our code is available at \href{https://github.com/yransun/LegoPET}{https://github.com/yransun/LegoPET}.

\subsection{Baselines}
We compared \techname{} against four baselines: DeepPET~\cite{haggstrom2019deeppet}, cGAN~\cite{mirza2014conditional}, Pix2Pix~\cite{isola2017image} and cDPM~\cite{saharia2022palette, ho2022classifier}. DeepPET is a regression-based method that utilizes a deep convolutional encoder-decoder network to map sinograms directly to PET images. cGAN is a standard conditional GAN for image reconstruction. Pix2Pix uses a conditional least squares GAN for image-to-image translation tasks. cDPM is a standard sinogram-conditioned diffusion model, which can also be regarded as ``\textit{\techname{} w/o guidance}". We used the publicly provided repository configurations for all baseline models and trained them until full convergence.

\subsection{Reconstruction Results}
\label{sec:recon_results}
We summarize the average pixel-wise PSNR/SSIM values across all reconstructed slices and the number of trainable parameters for each method in Table~\ref{numbers}. \techname{} demonstrates a 0.59dB improvement in PSNR compared to cDPM, validating the effectiveness of the proposed hierarchical feature-guided module. Additionally, we observe a substantial improvement of approximately 5dB/3.7dB in PSNR when compared with cGAN/Pix2Pix. Although DeepPET achieves comparable PSNR values to \techname{}, the PET images reconstructed by DeepPET are significantly less aligned with human visual perception, as shown in Fig.~\ref{fig:method}. This may be due to \techname{}'s reliance on diffusion-based generative models and stochastic posterior sampling, which compromises pixel-wise distortion (leading to lower PSNR) while maintaining fidelity to the target image~\cite{ren2023multiscale,ledig2017photo}. Visual reconstruction results are further illustrated in Fig.~\ref{fig:method}, with side-by-side comparisons of four baseline methods. \techname{} generates the most realistic reconstruction results among all methods, with the second lightest color in the squared error maps. cGAN, Pix2Pix, and cDPM introduce numerous artifacts, while DeepPET produces blurry reconstructions with unrealistic fine details.

\begin{table}[t!]
\centering
\setlength{\tabcolsep}{2.5pt} 
\renewcommand{\arraystretch}{1} 
\small 
\begin{tabular}{l|l|c|c|c}
\hline
\multirow{2}{*}{\textbf{Category}} & \multirow{2}{*}{\textbf{Method}} & \multicolumn{3}{c}{\textbf{Metrics}} \\ \cline{3-5}
                                   &                                  & PSNR $\uparrow$  & SSIM $\uparrow$  & MParam. \\
\hline
\multirow{1}{*}{Regression-based} 
& DeepPET \cite{haggstrom2019deeppet} &\textcolor{red}{\textbf{28.30}}&0.944& 11.02\\
\hline
\multirow{2}{*}{GAN-based} 
& cGAN \cite{mirza2014conditional} & 22.58 & 0.864 & 11.37 \\
& Pix2Pix \cite{isola2017image} & 23.91 & 0.922 & 11.37 \\
\hline
\multirow{2}{*}{Likelihood-based} 
& cDPM \cite{ho2022classifier}& 27.00 & \textcolor{blue}{\textbf{0.945}} & 35.71\\
& \textbf{\techname{} (Ours)} & \textcolor{blue}{\textbf{27.59}} & \textcolor{red}{\textbf{0.956}} & 35.71\\
\hline
\end{tabular}
\caption{\textbf{Quantitative Evaluation using PSNR, SSIM, and Number of Trainable Parameters (MParam).} Red and blue colors indicate the best and second-best results respectively.}
\label{numbers}
\end{table}

\subsection{Ablation Study}

\begin{figure}[t!]
    \centering
    \includegraphics[width=\linewidth]{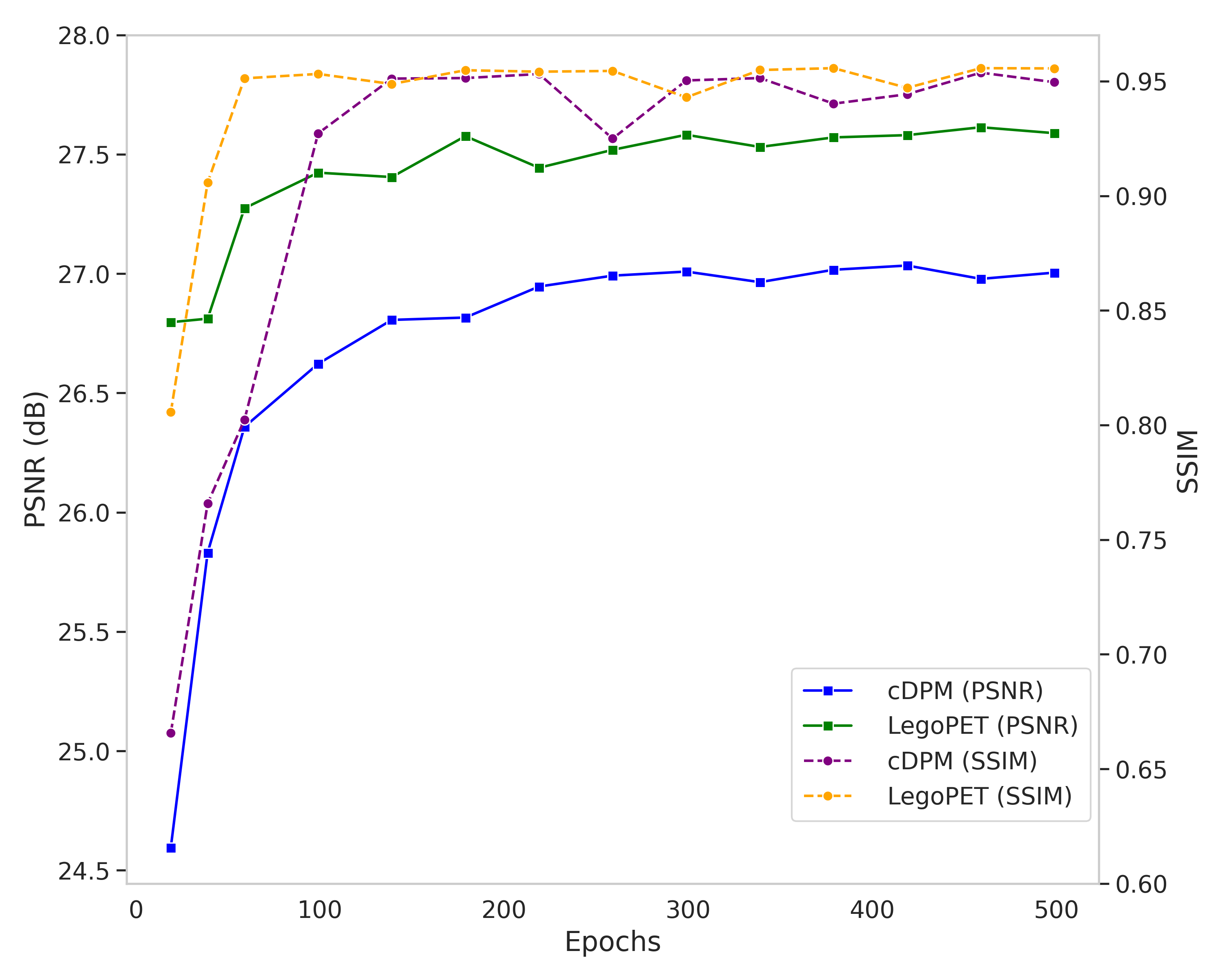}
    \caption{\textbf{Effectiveness of Hierarchical Feature Guidance.} We compare the performance of \techname{} and cDPM within 500 epochs in terms of PSNR and SSIM. We regard cDPM as ``\textit{\techname{} w/o guidance}", and \techname{} as ``\textit{cDPM w/ guidance}".}
\label{fig:plot}
\end{figure}

To further verify the effectiveness of our proposed \techname{}, we conducted an ablation study. For a fair comparison, we used the same training settings for both cDPM and \techname{}, including learning rate and the number of trainable parameters. Classifier-free diffusion guidance \cite{ho2022classifier} was disabled during the ablation study, i.e. $p_{dp}=0$. We compared their performance every 20 epochs in terms of PSNR/SSIM over a total of 500 epochs. The results presented in Fig.~\ref{fig:plot} reveal several key insights as expected. First, \techname{} consistently outperforms cDPM in terms of quantitative metrics, which underscores the benefits of integrating hierarchical feature representations with rich low- and high-frequency information from learned \denet{} into a cDPM. Second, \techname{} illustrates strong performance even at early epochs, suggesting its potential to improve training efficiency.

\section{Conclusion}
\label{sec:typestyle}

We propose \techname{}, which enables high-perceptual quality PET image reconstruction that preserves geometric structure and sharp edges from raw sinograms. The experimental results show that \techname{} outperforms several baseline algorithms in terms of visual inspection and PSNR/SSIM metrics. The ablation study further validates that our novel hierarchical feature-guided method has significant potential to improve both reconstructed image quality and training efficiency. Our future work will focus on extending current model into an efficient 3D version and validating on patient data.

{\small
\bibliographystyle{ieee_fullname}
\bibliography{egbib}
}

\end{document}